\documentclass[aps,amsmath,amssymb,twocolumn,prl,longbibliography]{revtex4-1}
\usepackage{graphicx} 
\usepackage{amsfonts,amsmath,amssymb}
\usepackage{amsthm}
\usepackage{dcolumn}
\usepackage{dsfont,bm}
\usepackage[colorlinks=true,linkcolor=blue,pagecolor=blue,filecolor=blue,menucolor=blue,urlcolor=blue,citecolor=blue,anchorcolor=blue]{hyperref}
\usepackage{xcolor}
\usepackage{soul} 
\usepackage{amsbsy}
\usepackage{float}

\usepackage{bm}

\begin{document}
\title{Anisotropic spin dynamics in antiferromagnets with a nonrelativistic spin splitting}
\author{Konstantin S. Denisov$^{1}$}
\author{Igor {\v{Z}}uti{\'c}$^{1}$}

\affiliation{$^{1}$Department of Physics, University at Buffalo, State University of New York, Buffalo, NY 14260, USA}

\begin{abstract} 
Antiferromagnets (AFM) with a nonrelativistic spin splitting (NSS) of electronic bands expand the range of spin-dependent phenomena and their applications. A crucial understanding for both of them pertains to the inherent spin dynamics. We demonstrate that the $d$-wave NSS gives rise to extremely anisotropic spin dynamics of carriers  driven by the relaxation induced by motional narrowing. In contrast to the no relaxation of the spin polarization along the Neel vector, the dynamics of perpendicular spin components greatly varies from a long relaxation for a weak NSS to the fast-decaying oscillations for a strong NSS, a regime relevant to AFM with the $d$-wave NSS. The extreme anisotropy of the spin relaxation is transformed by the external magnetic field, which triggers the relaxation of the parallel spin component and also suppresses all the relaxation rates at larger magnitudes. The predicted spin dynamics of the $d$-wave NSS, known also from altermagnets, can be used as their experimental fingerprints and guide future applications. 
\end{abstract}

\date{\today}
\maketitle

Spatially modulated 
magnetic fields can lead to a finite spin splitting of carriers in crystals even with no net magnetization~\cite{Pekar1964:JETP}. 
This mechanism has been recently revisited for antiferromagnets (AFM)~\cite{Smejkal2020:SA,Yuan2020:PRB,Hayami2020:PRB,Yuan2021:PRM,Mazin2021:PNAS,Egorov2021:JPCL,Smejkal2022:PRXb,Liu2022:PRX,Krempasky2024:Nature,Zhu2024:Nature}, 
where an exchange interaction between 
carriers and magnetic sublattices is site-dependent. A spin group analysis~\cite{Litvin1974:P,Litvin1977:ACA,
Hayami2019:JPSJ,Smejkal2020:SA,Yuan2020:PRB,Hayami2020:PRB,Yuan2021:PRM,Mazin2021:PNAS,Egorov2021:JPCL,Smejkal2022:PRXb,Liu2022:PRX,Chen2023enumeration,Jiang2023enumeration,Xiao2023spin}  
indicates the existence of 
material classes featuring AFM-induced spin splittings, including 
fully nonrelativistic NSS, also referred to as the altermagnetism~\cite{Smejkal2022:PRX,Mazin2022:PRX}.

This NSS arises in collinear AFM when sublattices with opposite magnetizations are connected only by crystal-rotation symmetries (but not by translations or spatial inversion)~\cite{Hayami2019:JPSJ,Smejkal2020:SA,Yuan2020:PRB,Hayami2020:PRB,Yuan2021:PRM,Mazin2021:PNAS,Egorov2021:JPCL,Smejkal2022:PRXb,Liu2022:PRX}, 
and it has a fixed spin quantization axis parallel to the Neel vector, $\bm{l}$, 
accompanied by a no net magnetization.
Vanishing NSS along high symmetry lines in the Brilliuon zone (BZ)
suggests the notation of the high-order magnetism~\cite{Smejkal2022:PRX}, 
e.g.~$d$-wave magnetism in analogy with the $d$-wave superconductivity. 
The magnitude of the $d$-wave NSS (meV to eV) 
is mainly determined by an atomic exchange interaction and can significantly exceed 
spin-orbital coupling (SOC) mechanisms. Such NSS can modify
superconducting~\cite{Ouassou2023:PRL,Zhu2023:PRB}, 
optical~\cite{Samanta2020:JAP}, topological~\cite{Bonbien2021:JPD,Fernandes2024:PRB,Antonenko2024:X} 
properties. 

\begin{figure}[b]
	\centering
	\includegraphics[width=.4\textwidth]{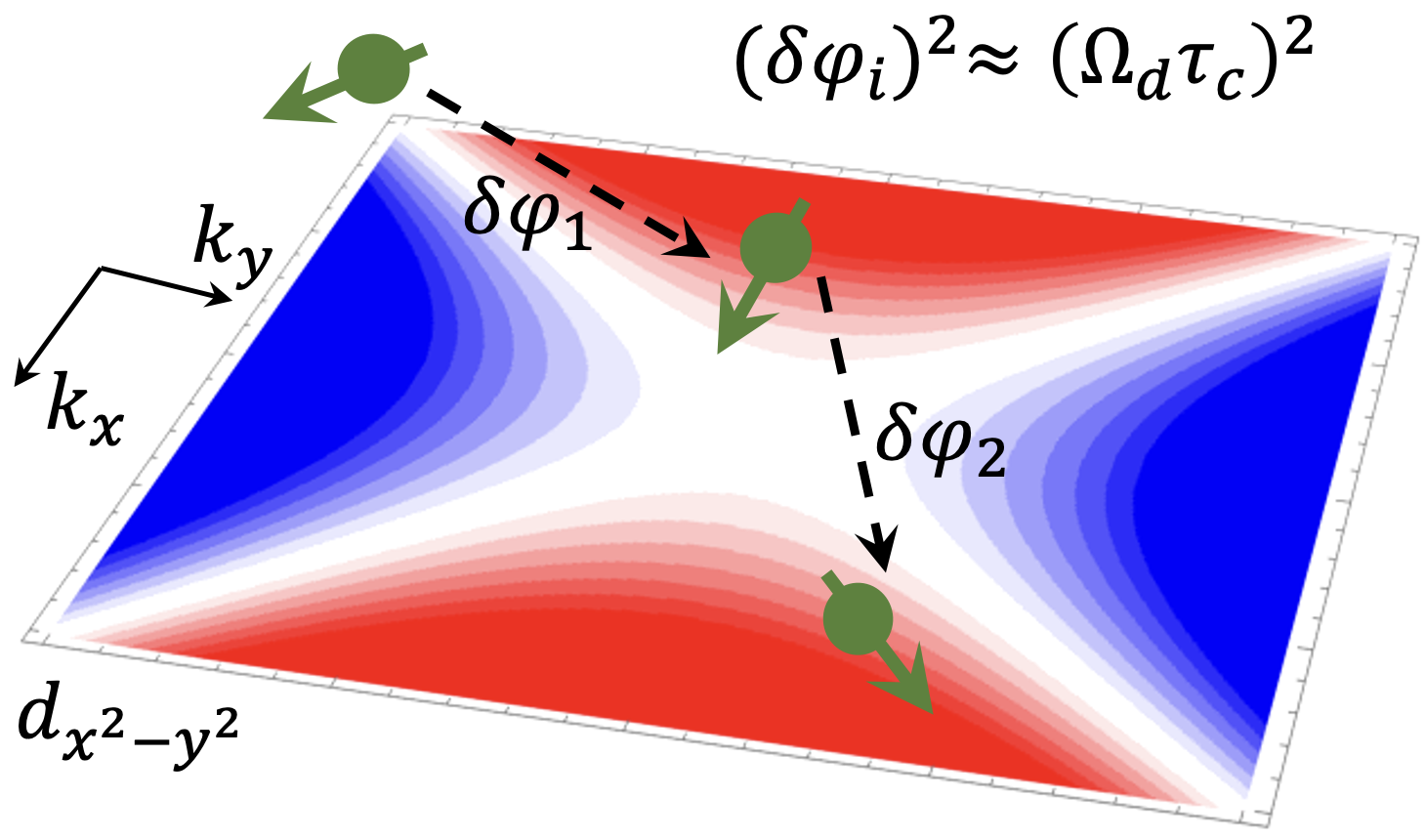}
	\caption{Dephasing, $\delta \varphi_i$, of the electron in-plane spin 
		due to its diffusion in the $d$-wave nonrelativistic spin splitting (red/blue for +/$-$ values) in AFM. 
		The Larmor frequency magnitude is $\Omega_{\rm d}$, 
		the correlation time, $\tau_c$, and the wave vector, ${\bm k}$.} 
	\label{fig:1}
\end{figure}

In this work, we demonstrate that 
the $d$-wave altermagnetism has a remarkable anisotropic spin dynamics 
of carriers, controlled by the in-plane 
magnetic field 
and sensitive to the interplay between the magnitude of NSS 
and disorder-induced relaxation from motional narrowing.~\cite{Dyakonov1972:SPSS,Dyakonov1986:SPS,Zutic2004:RMP,Slichter:2013}. 
This spin dynamics has a striking difference from the SOC-driven one in semiconductors~\cite{Zutic2004:RMP,Fabian2007:APS,Wu2010:PR,Glazov2010:PE,Averkiev2002:JPCM}.

Since NSS varies with the wave vector ${\bm k}$, it is equivalent to a $k$-dependent Larmor frequency $\bm{\Omega}_{\rm d}(\bm{k})$, 
leading to the precession of an electron spin, $\bm{s}$,  $\dot{\bm{s}} = \bm{\Omega}_{\rm d} \times \bm{s}$. 
For a diffusive motion, due to scattering on impurities or phonons, 
electron's momentum changes randomly, see Fig.~\ref{fig:1}. Between the collisions, 
due to NSS and its average amplitude  $\Omega_{\rm d}$, $\bm{s}$ rotates 
at a random angle, 
$\delta \varphi_i \approx \Omega_{\rm d} \tau_c$, 
 determined by a correlation time $\tau_c$, typically given by the momentum relaxation time, $\tau_p$, or
 the interaction time of electrons with phonons and holes~\cite{Zutic2004:RMP}. 

For a weak NSS, $\Omega_{\rm d} \tau_c \ll 1$, 
this rotation is interrupted by a momentum scattering 
and with a different $\bm{\Omega}_{\rm d}$, 
its randomness 
the clockwise and counterclockwise rotation are equally likely. 
The average angle does not change, but, as in the random walk of 
$t/\tau_c$ steps, the root-means-square is 
$\varphi_{\rm rns}^2(t) \approx \Omega_{\rm d}^2 \tau_c^2 (t/\tau_c)$
known from the motional narrowing and 
Dyakonov-Perel spin relaxation~\cite{Zutic2004:RMP,Dyakonov1972:SPSS,Dyakonov1986:SPS}. 
$\varphi_{\rm{rms}}(t)=1$, defines the spin-relaxation rate $\Omega_{\rm d}^2 \tau_c$ of the ensemble spin density, 
$\bm S$~\cite{Dyakonov1972:SPSS,Dyakonov1986:SPS}: the faster the momentum relaxation, the slower the spin relaxation.
We study this dynamics from
the Boltzmann kinetic equation for the density matrix~\cite{Dyakonov:2008},
$\rho_k = f_k \hat{I} + \bm{S}_k \bm{\sigma}$, 
where $f_k$ and $\bm{S}_k \bm{\sigma}$ 
are the $k$-resolved scalar and spin parts of the density matrix, while $\bm{\sigma}$ is the vector of Pauli matrices. 

The $d$-wave NSS ensures a fixed spin quantization axis
and a constant orientation of random field, $\bm{\Omega}_{\rm d}$, along $\bm{l}$. 
In a two-dimensional (2D) model,
the Hamiltonian is
\begin{equation}
	H = \frac{\hbar^2 \bm{k}^2}{2 m} + 
	\hbar\frac{\lambda}{2} \left( k_x^2 - k_y^2 \right) \sigma_z + \hbar\frac{1}{2} \bm{\Omega} \cdot \bm{\sigma},
	\label{eq:Ham0}
	\end{equation}
where $\hbar$ is the Plank constant, $m$ is a scalar effective mass, $\lambda$ is the NSS strength, defining 
$\bm{\Omega}_{\rm d} = \lambda k^2 \cos{2 \varphi} \hat{\bm{z}}$, with $\varphi$ is a polar angle of $\bm{k}$,
and the average value, $\Omega_{\rm d} = \lambda \langle k^2 \rangle^{1/2}$.
The last term in Eq.~(\ref{eq:Ham0}) accounts for a finite in-plane magnetization from an applied magnetic field, ${\bm B}$, with the Larmor frequency, $\bm{\Omega}\propto {\bm B}$,
and ${\bm s}=\bm{\sigma}/2$. 
We consider an easy-axis AFM, such that the application of the in-plane $\bm{B}$ tilts the sublattice magnetic moments to produce a finite in-plane magnetization. The exchange interaction between itinerant electrons and magnetized AFM enhances the magnitude of $\bm{\Omega}$.

\begin{figure}[t]
	\centering
	\includegraphics[width=.45\textwidth]{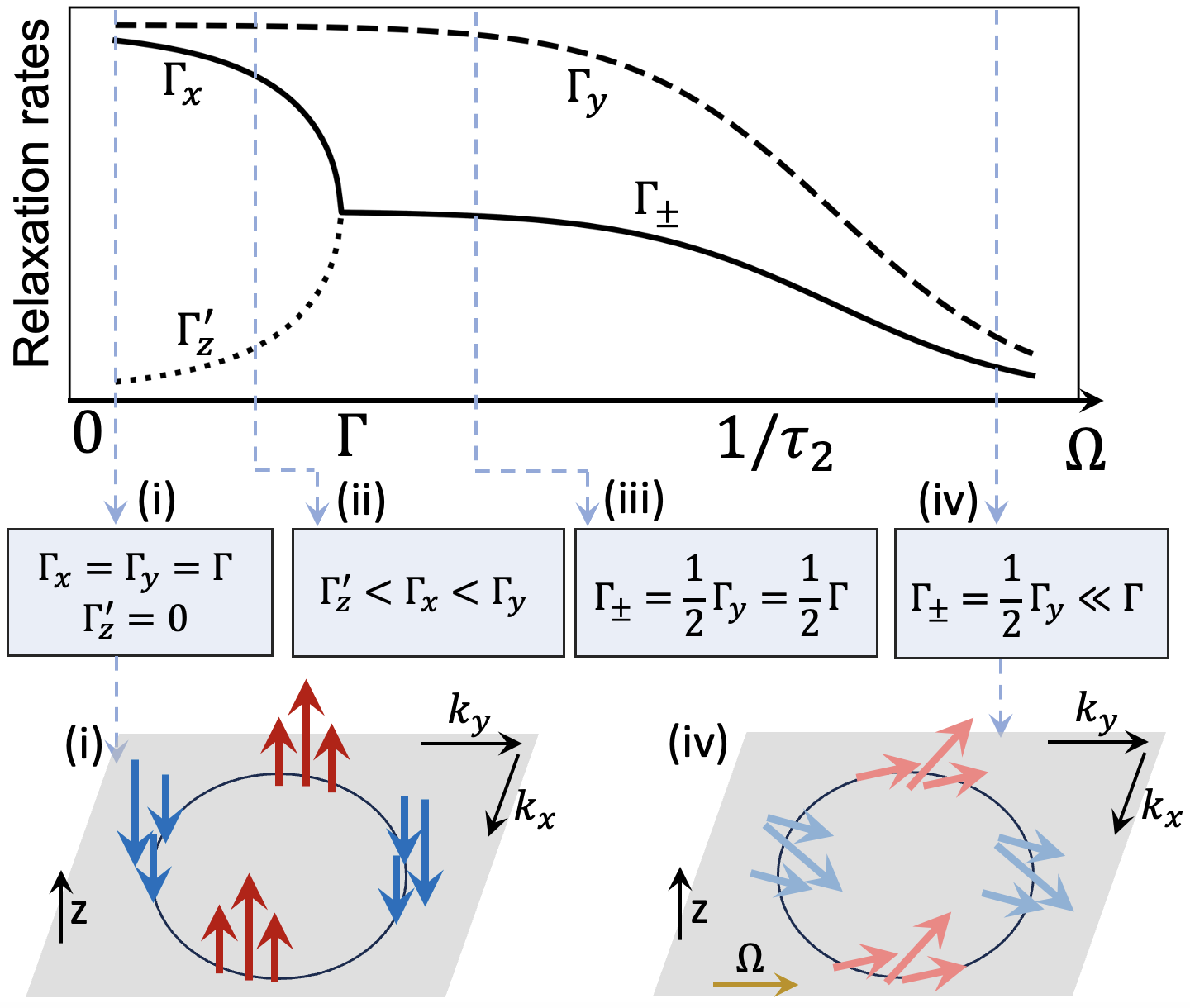}
	\caption{The evolution of the spin-relaxation rates with the Larmor frequency, $\Omega$, $\propto$ to the
	in-plane magnetic field $\parallel \hat{\bm{y}}$, shows distinct regimes at $\Omega_{\rm d} \tau_2 \ll 1$: (i)-(iv), labeled by their anisotropy 
	 ($1/\tau_2$ is for the second harmonic). 
	 The orientation of an effective magnetic field, defined by $\bm{\Omega}_{\rm d} + \bm{\Omega}$, 
	 changes from (i) out-of-plane to the in-plane (iv) configuration.}
	\label{fig:2}
\end{figure}

For a weak NSS, $\Omega_{\rm d} \tau_c \ll~1$, we identify different
spin-relaxation regimes (i)-(iv) as a function of 
an applied in-plane magnetic field $\sim \Omega$. In Fig.~\ref{fig:2}, these regimes are characterized by different
spin-relaxation rates, $\Gamma$.
(i) at $\Omega =0$ ($B=0$)  the spin-relaxation rate tensor~\cite{Dyakonov1972:SPSS,Dyakonov1986:SPS} is 
\begin{equation}
	\Gamma_{\alpha \beta} = 
	\langle 
	\left(\bm{\Omega}_{\rm d}^2 \delta_{\alpha \beta} -\Omega_{\rm d, \alpha} \Omega_{\rm d, \beta}\right) \tau_2
	\rangle,
\end{equation}
where $\alpha,\beta$ are Cartesian axes,  $\delta_{\alpha \beta}$ is the Kronecker symbol,
the brackets denote the averaging over angles and energies. 
$\tau_c$ is determined by the relaxation of the second angular 
harmonic of $\bm{S}_k$ with time $\tau_2$.
The relaxation of the in-plane spin components, $S_{x,y}$ is isotropic with rates
\begin{equation}
	\Gamma_x = \Gamma_y \equiv \Gamma = 
	(1/2) \Omega_{\rm d}^2 \tau_2.
\end{equation}
The relaxation of $S_z$, parallel to $\bm{l}$, 
is completely suppressed, $\Gamma_z=0$, since
$\bm{\Omega}_{\rm d} \parallel \hat{\bm{z}}$ for every $\bm{k}$. 
Therefore, in Fig.~\ref{fig:2}, (i) is characterized by a giant out-of-plane anisotropy
and no spin relaxation parallel to $\bm{l}$. 

By adding a small in-plane magnetization,
$\bm{\Omega} = \Omega \hat{\bm{y}}$, in Eq.~(\ref{eq:Ham0}), 
the spin dynamics is described by the precession 
\begin{equation}
		\label{eq:precess}
	\dot{\bm{S}} = \bm{\Omega} \times \bm{S} - \tilde{\Gamma} \bm{S},
\end{equation}
$\tilde{\Gamma} = {\rm diag}(\Gamma_x, \Gamma_y, 0)$. 
For $\Omega \ll \Gamma$, 
$S_z$ changes slowly compared to $S_{x,y}$, 
with $\Omega$ establishing $S_{x} \approx (\Omega / \Gamma) S_z$. The slow dynamics of $S_z$ is found from $\dot{S}_z \approx - \Gamma_z^\prime S_z$~with 
\begin{equation}
	\label{eq:GamZ}
\Gamma^\prime_z = \Omega^2 / \Gamma = 
		2 \Omega^2/(\Omega_{\rm d}^2 \tau_2),
	\end{equation}
where $\Gamma^\prime_z \ll \Gamma$ is the effective spin-relaxation rate.
In this regime, $\Gamma_{x,y}$ and $\Gamma_z^\prime$ 
exhibit opposite behavior with respect to $\tau_2$. While $\Gamma_{x,y} \propto \tau_2$, 
as expected for 
motional narrowing suppressed by disorder,
 $\Gamma^\prime_z \propto 1/\tau_2$
resembles more the Elliot-Yafet spin-relaxation mechanism~\cite{Elliott1954:PR,Yafet:1963,Zutic2004:RMP} 
and it is enhanced at shorter $\tau_2$.

(ii) By increasing $\Omega$ in the region $\Omega \lesssim \Gamma$,   
the in-plane spin-relaxation becomes also anisotropic, 
$\Gamma^\prime_z < \Gamma_x < \Gamma_y$, see Fig.~\ref{fig:2}. 
While the relaxation rate $\Gamma_y = \Gamma$ remains unchanged, the dynamics of coupled $S_x$ and $S_z$ 
follows the eigenfrequencies of Eq.~(\ref{eq:precess})
\begin{equation}
	\omega_{\pm} = (i/2)\left( - \Gamma \pm \sqrt{\Gamma^2 - 4 \Omega^2} \right). 
\end{equation}
The component of  the $\omega_+$ ($\omega_-$)
mode is mostly concentrated in $S_x$ ($S_z$), with the relaxation rate
$\Gamma_+ = - {\rm Im}[\omega_+] \approx \Gamma_x$ 
($\Gamma_- = - {\rm Im}[\omega_-] \approx \Gamma_z^\prime$).

(iii) At a larger magnetic field, $\Omega \gtrsim \Gamma$,  
the decaying dynamics of $S_{x,z}$ acquires the oscillatory character. 
$\omega_{\pm}$ gains nonzero real parts corresponding to the oscillation frequencies. 
The relaxation becomes isotropic in the $xz$-plane, perpendicular to $\bm{\Omega}$, with the damping rates, $\Gamma_\pm = \Gamma_y / 2$,  where $\Gamma_y = \Gamma$. 
From (i) to (iii)  the spin-relaxation anisotropy is transformed 
from the giant out-of-plane to the in-plane configuration following the direction of $\bm{\Omega}$. 

(iv) Finally, at an even stronger $\Omega \tau_2 \gg 1$,
the spin relaxation is suppressed.  $\bm{\Omega}$ quenches 
the spin precession in $\bm{\Omega}_{\rm d}$ and reduces a random angle acquired by 
the electron spin between subsequent collisions, suppressing the relaxation as well studied 
in semiconductors~\cite{Ivchenko1973:SPSS,Bronold2002:PRB,Glazov2004:PRB,Glazov2007:SSC,Griesbeck2009:PRB,Offidani2018:PRB}. 
For the considered geometry, $\bm{\Omega} = \Omega \hat{\bm{y}}$, 
the longitudinal, $\Gamma_y$, and transversal $\Gamma_{\pm}$ damping rates are~\cite{Fabian2007:APS}
\begin{equation}
	\Gamma_y = \frac{ \langle ({\Omega_{{\rm d},x}^2} + {\Omega_{{\rm d},z}^2}) \tau_c \rangle }{1 + \Omega^2 \tau_c^2 }, 
	\quad
	\Gamma_{\pm} = \langle {\Omega_{{\rm d},y}^2} \tau_c \rangle  + 	(1/2) \Gamma_y. 
	\notag
\end{equation}
The $d$-wave NSS has only a single component 
$\bm{\Omega}_{\rm d} \parallel \hat{\bm{z}}$ 
(with $\Omega_{\rm d,x}=\Omega_{\rm d,y}=0$). 
The relaxation rates 
\begin{equation}
\Gamma_y=  \Omega_{\rm d}^2 \tau_2/ 2 [1 + \left( \Omega \tau_2 \right)^2], 
\qquad 	\Gamma_{\pm} =  \Gamma_y/2, 
\end{equation}
are suppressed for $\Omega \tau_2 \gg 1$, 
$\Gamma_y \approx \Omega_{\rm d}^2 / (2 \Omega^2 \tau_2) \ll \Gamma$ 
and 
retain the uniaxial anisotropy $\Gamma_x = \Gamma_z = \Gamma_y/2$. 
This is unlike the SOC-driven spin relaxation in 2D structures, where
the in-plane ${\bm B}$ does not suppress $\Gamma_\pm$~\cite{Duckheim2006:NP,Rakitskii2024:P}, 
since SOC-driven
random field has a nonzero $y$ component.

Our prior analysis focused on a disordered regime, $\Omega_{\rm rnd} \tau_c \ll 1$, when the random precession frequency, $\Omega_{\rm rnd}$,  
was small compared to the scattering frequency. 
This condition is typically fulfilled for conduction band electrons in semiconductors with  SOC-driven spin relaxation~\cite{Wu2010:PR,Glazov2010:PE,Averkiev2002:JPCM}. 
Realizing  $\Omega_{\rm rnd} \tau_c \gtrsim 1$  transforms a slow exponential spin decay  to  
rapidly decaying oscillations~\cite{Ferreira1991:PRB,Gridnev2001:JETPL,Leyland2007:PRBb,Leyland2007:PRB,Liu2011:PRB,Cummings2017:PRL,Szolnoki2017:PRB,
Offidani2018:PRB}.
In semiconductors that requires either ultra-high electron mobilities with 
$\tau_p \rightarrow~$ps ~\cite{Leyland2007:PRBb,Leyland2007:PRB}, 
or strong SOC in a meV range~\cite{Cummings2017:PRL,Benitez2018:NP,Offidani2018:PRB,Zutic2019:MT,Rakitskii2024:P}.

Since $d$-wave NSS                                                                                             arises from an exchange interaction in AFM and does not have a relativistic origin, 
 $\Omega_{\rm d} \tau_2 \gtrsim 1$ can be already realized for moderate electron mobilities. 
Taking $\hbar \Omega_{\rm d} \approx 2$~meV,  consistent with a lower estimate of NSS for MnF$_2$ 
in the conduction band bottom~\cite{Yuan2020:PRB,Yuan2021:PRM}, 
$\Omega_{\rm d} \tau_2 > 1$ already at $\tau_2 \gtrsim 0.5$ ps. 
Remarkably, in this case, 
the oscillating dynamics of $S_{x,y}$ 
coexists with a completely undamped $S_z$, parallel to $\bm{l}$. 
The dynamics of $S_z$ will be activated only by applying in-plane $\bm{\Omega}$. 

To investigate the spin dynamics for an arbitrary $\Omega_{\rm d} \tau_2$, we use
the previously mentioned  Boltzmann kinetic equation for the density matrix.   
For the 2D 
electron gas, $\bm{S} = {\rm Tr}[ \rho \hat{\bm{s}} ] = \sum_k \bm{S}_k$, 
where  ${\rm Tr}$ is the trace.
$\bm{S}_k(t)$ can be found from the kinetic equation~\cite{Dyakonov:2008}
\begin{align}
	&		\frac{ \partial \bm{S}_k }{\partial t} = 
	\left(  \bm{\Omega}_{\rm d} + \bm{\Omega} \right) \times \bm{S}_k + \mathcal{I}[\bm{S}_k],
\end{align}
where we assume that the collision integral, $\mathcal{I}$, depends only on $\bm{S}_k$,  
($f_k$ and $\bm{S}_k \bm{\sigma}$ are decoupled), and 
that the relaxation, due to electron elastic scattering on nonmagnetic scalar impurities, 
is treated in the relaxation time approximation. 
We expand $\bm{S}_k$ in a series over angular harmonics, $\bm{S}_n = \bm{S}_{n}^{+} \cos{n \varphi}  	+ \bm{S}_{n}^{-}  \sin{n \varphi}$, with
\begin{equation}
	\bm{S}_k=  \bm{S}_0 + \sum_{n \ge 1}  \bm{S}_n,
\end{equation}
and express $\mathcal{I}$ in terms of $\tau_n$, the 
relaxation time of $n^{th}$ harmonic
\begin{equation}
	\label{eq:tau-aprx}
\mathcal{I} = - \sum_{n \ge 1} \frac{\bm{S}_n}{\tau_n}, 
\quad
	\frac{1}{\tau_n} =  \frac{2 \pi}{\hbar} n_i g_0 \Big\langle  u^2(\theta) (1- \cos{n \theta}) \Big\rangle,
\end{equation}
here $u(\theta) = u_{kk'}$ 
is the matrix element of an impurity potential, $u$,
$\theta = \varphi' - \varphi$ is the scattering angle 
from an incident $\bm{k}$ to a final $\bm{k}'$ state, 
angular brackets denote the $\theta$ averaging, 
$n_i$ is the sheet density of impurities, 
$g_0 = m /2 \pi \hbar^2$ is an effective density of states. 
We assume $2m \lambda  \ll 1$, 
and that the kinetic energy, $E \gg \hbar \Omega_{\rm d}$, so 
we can use angle-independent $g_0$ in Eq.~(\ref{eq:tau-aprx}). 
In this approximation, 
 $\bm{S}_k = \bm{S}(k,\varphi)$ 
depends on $\varphi$ and on $k = \sqrt{2 m / E}$ independently. 
In what follows, we consider degenerate 2D electron gas,
and $\bm{S}_k = \bm{S}(k_F,\varphi)$ 
with the Fermi wave vector $k_F = \sqrt{2 m \mu}$,  
where $\mu$ is the chemical potential, 
and we express $\bm{S} \approx g_0 \bm{S}_0$ and $\Omega_{\rm d} = \lambda k_F^2$. 

\begin{figure}[t]
	\centering
	\includegraphics[width=.45\textwidth]{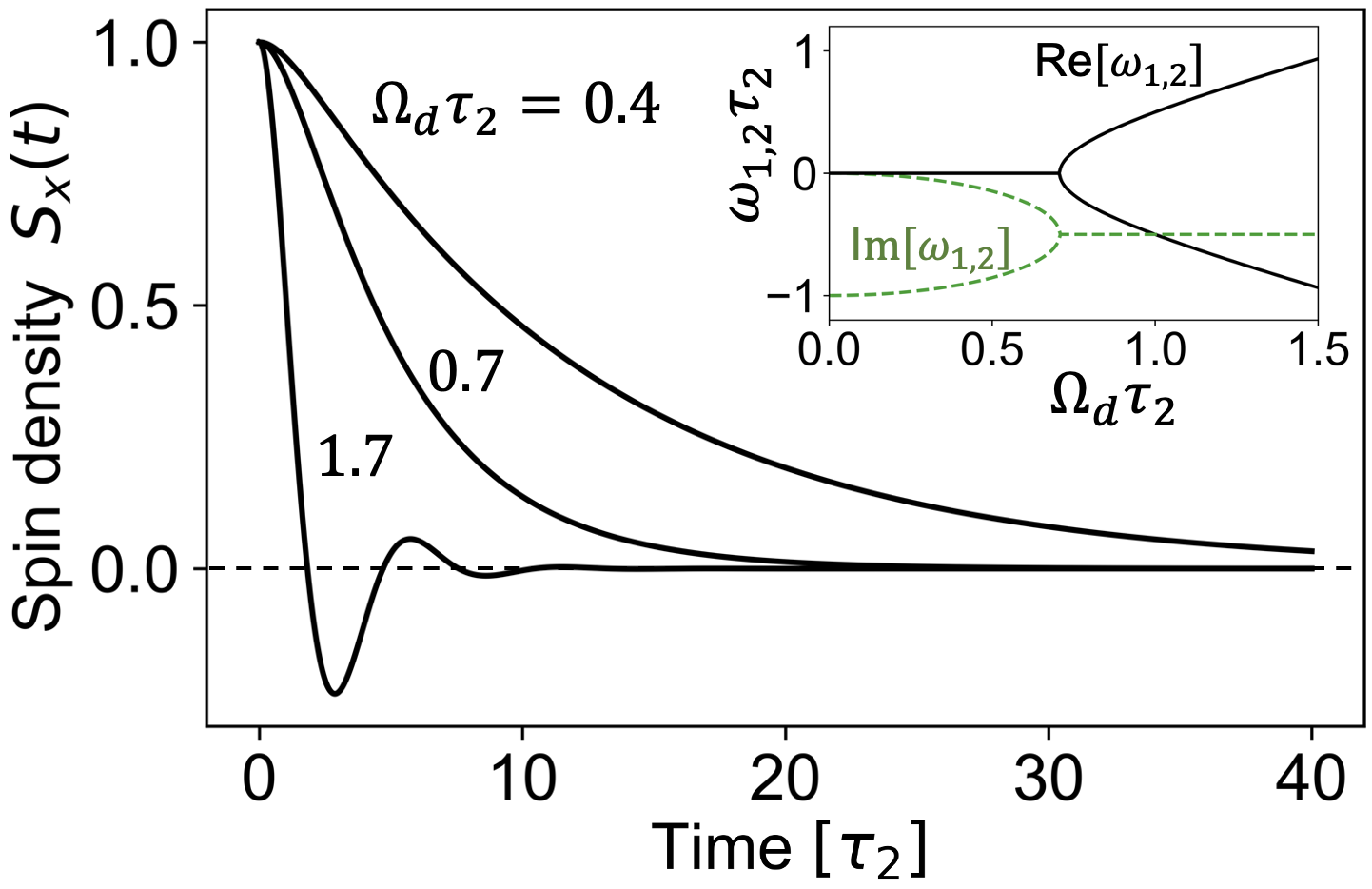}
	\caption{Dynamics of the in-plane spin density, $S_x(t)$, after an instantaneous excitation, calculated from Eq.~(\ref{eq:chi1}). 
	$\tau_2$ is the second harmonic relaxation time.  
  	$S_x(t)$ evolves from a slow decay at $\Omega_{\rm d} \tau_2 = 0.4$ to a 
	rapidly oscillating decay at $\Omega_{\rm d} \tau_2 = 1.7$ 
	The inset shows the trajectory of the susceptibility poles, $\omega_{1,2}$, obtained from Eqs.~(\ref{eq:chi1}).	
}
	\label{fig:3}
\end{figure}

To analyze the spin dynamics, we use Fourier components, $\bm{S}_k e^{- i \omega t}$, and study the trajectories of eigenfrequencies, determined from the equation
\begin{equation}
	\label{eq:freq_kinet}
		- i \omega \bm{S}_k = \left(\bm{\Omega}_{\rm d} + \bm{\Omega}\right) \times \bm{S}_k 
	- \sum_{n \ge 1} \frac{\bm{S}_n}{\tau_n}. 
\end{equation}
For $2m \lambda  \ll 1$, 
$\mathcal{I}$ does not contain~$\bm{S}_0$, 
and the relaxation of $\bm{S}_0$ and $\bm{S} = g_0 \bm{S}_0$ 
is driven by its coupling with higher angular harmonics. 
The term $\bm{\Omega}_{\rm d} \times \bm{S}_k = \Omega_{\rm d} \cos{2\varphi}(\hat{\bm{z}} \times \bm{S}_k)$ 
mixes $\bm{S}_0$ with even harmonics $n=2,4,\dots$. 
At $\Omega =0$, 
$S_{x}$ and $S_y$ are decoupled, 
forming the two 
independent series $(S_{0x},S_{2y},S_{4x},\dots)$ and $(S_{0y},S_{2x},S_{4y},\dots)$, we denote as $(x,y)$-series, 
displaying the same dynamics. 

In the biharmonic approximation (BHA) 
we only keep the two lowest harmonics ($n=0,2$).
For the $x$-series, Eq.~(\ref{eq:freq_kinet}) 
at $\Omega=0$ is reduced to $2\times 2$ matrix for $(S_{0x}, S_{2y})$  
with the eigenfrequencies given by 
\begin{equation}
	\label{eq:ome12}
	\omega_{1,2} = \frac{i}{2 \tau_2} \left( - 1 \pm \sqrt{1 - 2 (\Omega_{\rm d} \tau_2)^2} \right).
\end{equation}
In the disordered regime, $\Omega_{\rm d} \tau_2 \ll 1$, 
the $\omega_1$ mode describes the spin relaxation, 
$\omega_1 \approx - i\Gamma = - i \Omega_{\rm d}^2 \tau_2/2$, 
with $S_{0x} \gg S_{2y}$, 
and the $\omega_2$ mode corresponds to 
the rapidly decaying oscillations $S_{2y} \gg S_{0x}$. 
By increasing $\Omega_{\rm d}$, 
the components 
$S_{0x}, S_{2y}$ mix more effectively, resulting in 
the oscillatory dynamics when $\Omega_{\rm d} \tau_2 > 1/2$. 
For a weak disorder, 
$\Omega_{\rm d} \tau_2 \gg 1$, 
we get spin beatings with frequency $\Omega_{\rm d}/\sqrt{2}$ 
decaying with $\Gamma_{x,y} = 1/2 \tau_2$. 

The transverse spin susceptibility, $\chi(\omega)$, describing a linear spin response 
$S_x = \chi(\omega) \mathcal{T}_{x}$ 
to an applied spin-torque, $\bm{\mathcal{T}} = \mathcal{T}_x \hat{\bm{x}}$, 
can be obtained by adding $\bm{\mathcal{T}}$, to the right side of Eq.~(\ref{eq:freq_kinet}),
and for the BHA is given by
\begin{equation}
	\label{eq:chi1}
	\chi(\omega) = \frac{i \omega - 1/\tau_2}{(\omega - \omega_1)(\omega-\omega_2)},
	\qquad
\Omega_{\rm d} \tau_{4,6,\dots} \ll 1. 
\end{equation}
The corresponding evolution of $S_x(t)$ after an instantaneous initial excitation, $\mathcal{T}_x \propto \delta(t)$,  
is shown in Fig.~\ref{fig:3}. 
$S_x(t)$  is transformed from slow relaxation at $\Omega_{\rm d} \tau_2 \ll 1$ 
to the rapidly decaying oscillations for $\Omega_{\rm d} \tau_2 \gtrsim 1$. 

With a finite $\Omega$, there is an admixture of $S_{0z}$ to $S_{0x}, S_{2y}$ ($x$-series), 
and $S_{2z}$ to $S_{0y}, S_{2x}$ ($y$-series). 
A detailed $\Omega$-dependence of the damping rates, 
obtained from Eq.~(\ref{eq:freq_kinet}) in the BHA for $\Omega_{\rm d} \tau_2 \ll 1$, 
was discussed above, see Fig.~\ref{fig:2}. 
Here we provide $\omega_{\pm}, \omega_y$, related to $S_{x,z}$ and $S_y$ modes and 
valid for an arbitrary~$\Omega \tau_2$ 
in (iii) and (iv),  
$\Omega_{\rm d} \ll \Omega, 1/\tau_2$,
\begin{equation}
	\label{eq:om-om}
	\omega_\pm = \pm \Omega + \frac{i}{4} \frac{\Omega_{\rm d}^2 \tau_2}{1 \pm i \Omega \tau_2},	
	\qquad \omega_y = 
	\frac{i}{2} \frac{\Omega_{\rm d}^2 \tau_2}{1 + (\Omega \tau_2)^2}.
	\end{equation}
In addition to the suppression of the transverse spin relaxation, 
$\Gamma_{\pm} = -{\rm Im}[\omega_{\pm}] \approx \Omega_{\rm d}^2 / (4 \Omega^2 \tau_2) \ll \Gamma$, 
there is also a shift in its precession frequency, 
$\delta \omega_\pm \approx \Omega_{\rm d}^2 / 4 \Omega$.

\begin{figure}[t]
	\centering
	\includegraphics[width=.45\textwidth]{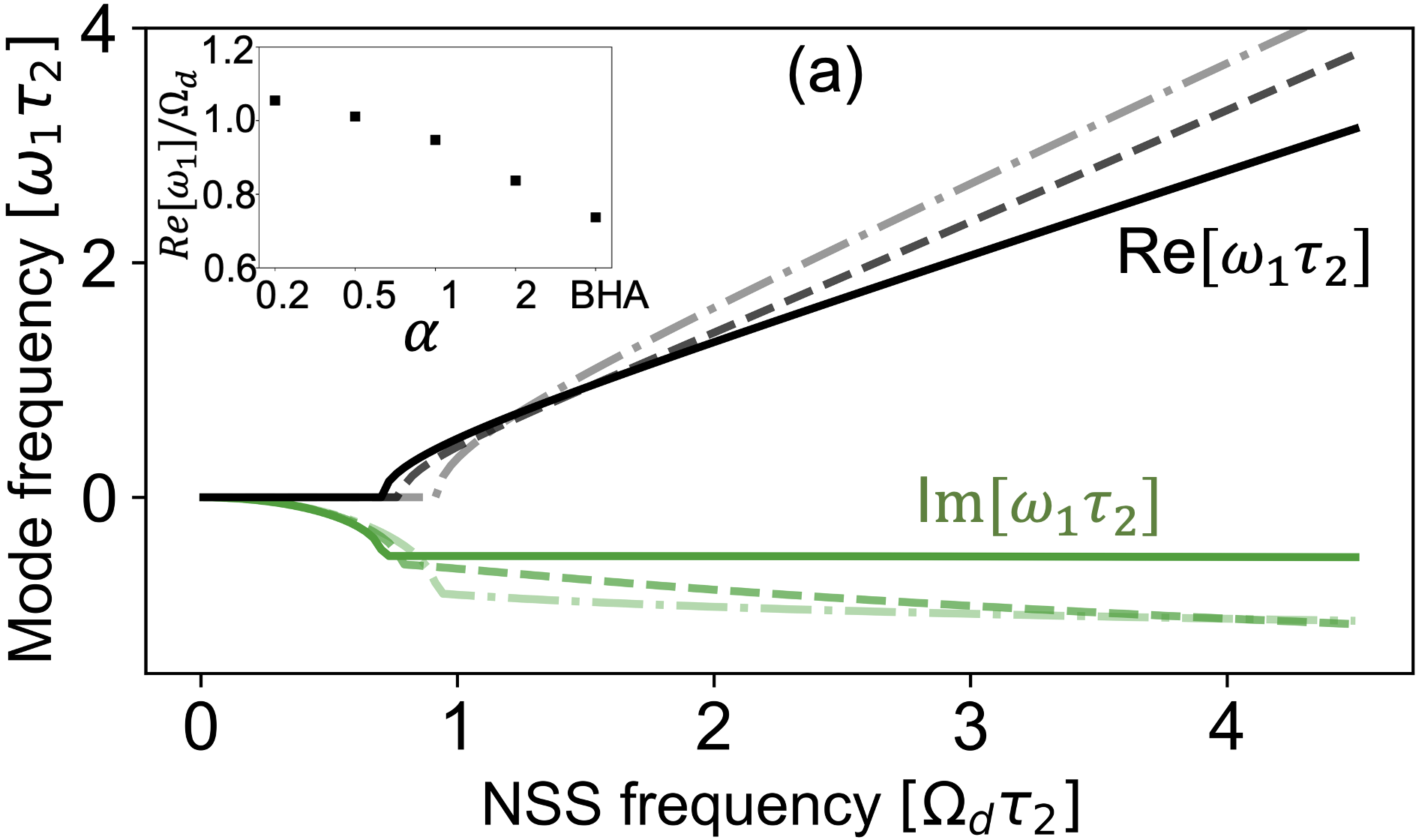}
		\includegraphics[width=.45\textwidth]{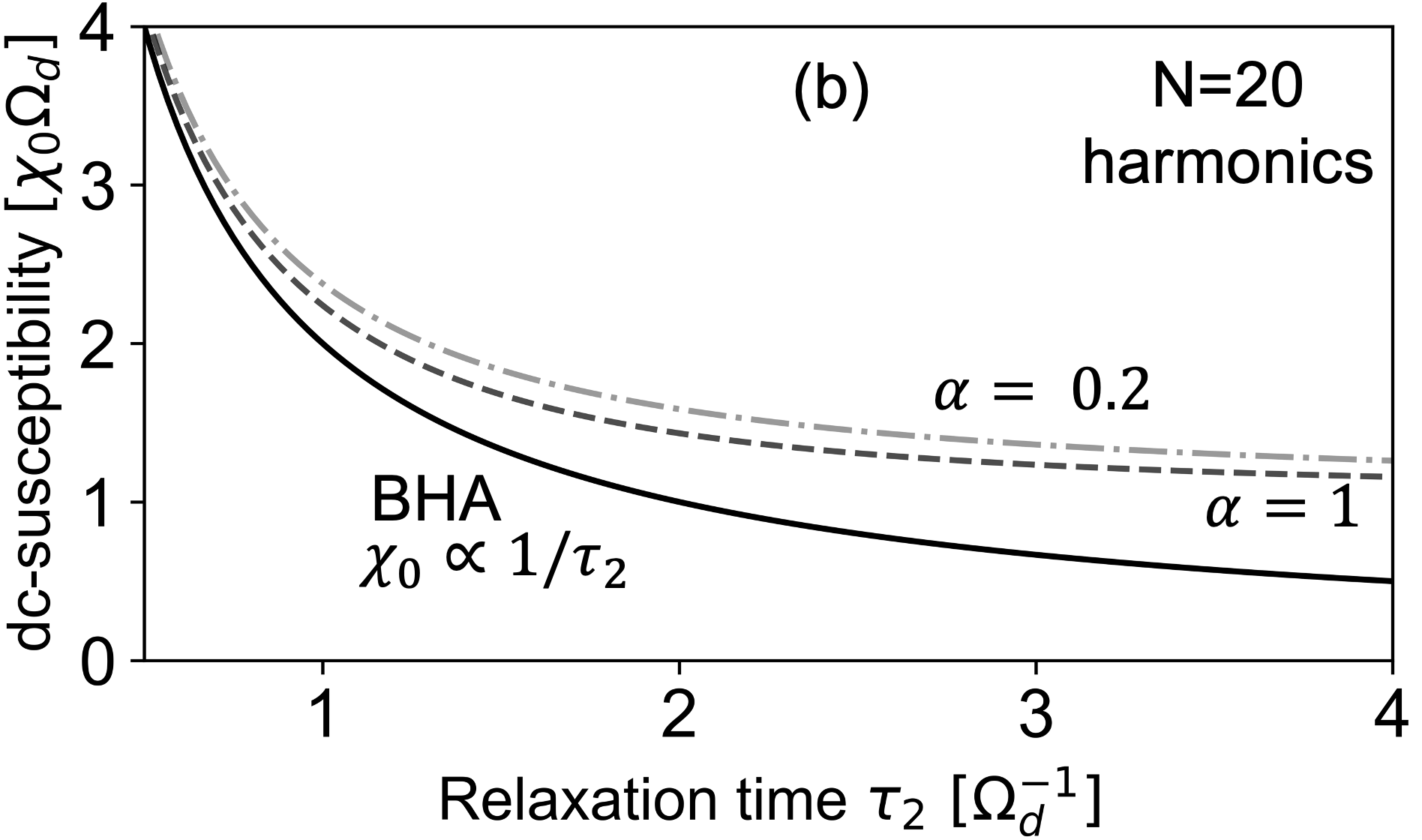}
	\caption{(a) Oscillation frequencies (black) and relaxation rates (green) of the $S_{0x}$-like mode as a function of $\Omega_{\rm d}$ 		
		for different models of the relaxation times ($\tau_{2n} = n^{-\alpha} 
	 \tau_2$) with $\alpha \gg 1$ (BHA, solid lines),  $\alpha = 1$ (dashed) and $\alpha = 0.2$ (dash-dotted). The inset shows the slope ${\rm Re}[\omega_1] / \Omega_{\rm d}$ for different $\alpha$.
        (b) The dependence of the susceptibility, $\chi_0(\tau_2)$, and $\Gamma_z^\prime = \chi_0 \Omega^2$.  For $\tau_2 \to \infty$, $\chi_0 \to 0$ in BHA (solid), 
        but $\chi_0 \approx \Omega_{\rm d}^{-1}$ for $N=20$ harmonics, $\alpha = 1$ (dashed) and $\alpha = 0.2$ (dash-dotted).}
	\label{fig:4}
\end{figure}

To examine the accuracy of BHA, we note that
it is realized when  $\Omega_{\rm d} \tau_{4,6,\dots} \ll 1$ 
and $\bm{S}_{4,6,\dots}$ are decoupled from $\bm{S}_{0,2}$. 
The condition $\tau_{4,6,\dots} \ll \tau_2$ 
implies a strongly anisotropic electron scattering in Eq.~(\ref{eq:tau-aprx}). 
For an isotropic-like scattering, the decoupling will be incomplete and will change 
the spin dynamics at $\Omega_{\rm d} \tau_2 \gtrsim 1$. 
For the regime (i) and $\Omega=0$, the trajectories of the eigenfrequency related to  $\bm{S}_0$  
are shown in Fig.~\ref{fig:4}(a) 
for different relaxation mechanisms (different types of disorder)
modeled by $\tau_{2n} \propto n^{-\alpha}$,~$\alpha > 0$. 
By changing $\alpha$ from $\alpha \gg 1$ (anisotropic)
to $\alpha=0.2$ (isotropic scattering), 
the damping rate changes from $\tau_2/2$ to $\tau_2$,  
while the inset of Fig.~\ref{fig:4}(a)
reveals that the magnitude of the beating frequency changes 
from $\Omega_{\rm d} / \sqrt{2}$ to $\Omega_{\rm d}$.

An extreme anisotropy of the spin dynamics and its tunability by external magnetic field remains attainable even at $\Omega_d \tau_2 \gtrsim 1$.
The relaxation of $S_{z}$ 
in the regime (ii)
is governed by the static limit of $\chi_0 = \chi(\omega \to 0)$. 
At $\Omega \ll \Gamma$ the dynamics of $S_z$ is slow 
compared to the in-plane components, 
and $S_{x}$ is induced by 
an effective spin torque, $\mathcal{T} = (\bm{\Omega} \times S_z \hat{\bm{z}})_x = - \Omega S_{0z}$, leading to the steady state, $S_{0x} = \chi_0 \mathcal{T}$.
With $S_{0x} \neq 0$, the spin torque $(\bm{\Omega} \times S_{0x} \hat{\bm{x}})_z$ 
is responsible for the relaxation  of $S_{0z}$ with the rate  
$\Gamma_z^\prime = \chi_0 \Omega^2$. 
While this is valid for an arbitrary $\Omega_{\rm d} \tau_2$, 
using Eq.~(\ref{eq:GamZ}) and $\Gamma_z^\prime = 2 \Omega^2 / \Omega_{\rm d}^2 \tau_2$ 
in the BHA at $\Omega_{\rm d} \tau_2 \gg 1$ is not accurate. 
This estimate suggests suppressing 
$\Gamma_z^\prime \propto 1/\tau_2 \ll 1$ with $\tau_2$. 
However, an accurate analysis involving higher angular harmonics 
(beyond BHA)
reveals that there is no suppression of $\Gamma_z^\prime$ 
even for $\tau_2  \rightarrow \infty$. Instead, $\Gamma_z^\prime$ reaches a constant, relaxation-independent value 
$\Gamma_z^\prime = \Omega^2 / \Omega_{\rm d}$. $\chi_0(\tau_2)$ is illustrated in Fig.~\ref{fig:4}(b) for different 
relaxation mechanisms (different $\alpha$).  
We calculate $\chi_0 = (\mathcal{M})_{0x,0,x}^{-1}$, where 
$\mathcal{M}$ is the matrix operator in Eq.~(\ref{eq:freq_kinet}) at $\omega=0$ for  $N=20$ harmonics.
From Fig.~\ref{fig:4}(b) we find  a universal value 
$\Gamma_z^\prime \approx \Omega^2 / \Omega_{\rm d}$  
at $\Omega_{\rm d} \tau_n \gtrsim 2$, 
an independent of the type of disorder.

When, at a large $\Omega_{\rm d}$,  
the higher harmonics $\bm{S}_{2,4,\dots}$  do not decay,  
the analysis cannot be reduced to a single spin-like mode, $\bm{S}_0$. 
For an isotropic scattering, 
characterized by a single relaxation time, $\tau_n = \tau$,
this occurs at $\Omega_{\rm d} \tau \gg 1$.  
To clarify this regime we turn to the spin dynamics in the clean limit. 
An electron with its spin, $\bm{s}_k$, 
has the spin torque 
$\bm{\Omega}_{\rm d} \times  \bm{s}_k$ 
and precesses in the $xy$-plane with $\Omega_{\rm d} \cos{2\varphi}$. 
Taking $\bm{S}(t=0) = S_0 \hat{\bm{x}}$, 
electrons with different $\bm{k}$ will precess at different frequencies, 
resulting in the dephasing of $\bm{S}$, 
without any scattering and due to the variation in $\Omega_{\rm d}$. 
$S_x(t)$ can be found by summing 
$\bm{s}_k(t)$ 
and, at $2 m \lambda \ll 1$, is given by
the Bessel function,~$J_0$
\begin{equation}
	S_x = S_0 \Big\langle \cos{\left(\Omega_{\rm d} t \cos{2\varphi} \right)} \Big\rangle_{\varphi} = S_0 J_0( \Omega_{\rm d} t). 
\end{equation}
At times $\Omega_{\rm d} t \gtrsim 1$, 
$S_x(t) \approx S_0 \sqrt{2 / \pi \Omega_{\rm d} t} \cos{\left( \Omega_{\rm d} t - \pi/4 \right)}$
precesses with $\Omega_{\rm d}$ and has a non-exponential decay. 
$\chi(\omega)$ is obtained by summing over contributing poles in  $\bm{s}_k(t)$ 
\begin{equation}
	\label{eq:sus_cl}
	\chi(\omega) = [\Omega_{\rm d}^2-(\omega+i/\tau)^2]^{-1/2},
	\qquad
	\Omega_{\rm d} \tau \gg 1, 
\end{equation}
and it cannot be reduced to a single pole contribution,  as in Eq.~(\ref{eq:chi1}) for BHA 
or for vanishing $\Omega_{\rm d} \tau_n \ll 1$ beyond BHA, 
since it features cuts as a function of $\omega$. 
In the regime (ii), $\Gamma_z^\prime = \chi_0 \Omega^2 = \Omega^2 / \Omega_{\rm d}$, obtained from the clean limit of Eq.~(\ref{eq:sus_cl}), 
is consistent with our analysis in Fig.~\ref{fig:4}(b), 
implying that $\chi_0$ and, therefore, $\Gamma_z^\prime$ at $\Omega_{\rm d} \tau \gtrsim 1$ are mostly related to the dephasing decay of $S_{x,y}$.

A rapidly changing landscape of altermagnets resembles the research in dilute magnetic semiconductors (DMS)
from several decades ago~\cite{Zutic2004:RMP,Timm:2019} 
where a growing family of their 
candidates and important phenomena, later discovered in other materials, were also accompanied by cautionary examples 
of extrinsic effects and alternative explanations. 
In this regard, the predicted zero magnetization of altermagnets 
could be fragile in the presence of SOC and, instead, transform 
them into weak ferromagnets with a nonvanishing magnetization~\cite{Milivojevic2024:2DM,Maznichenko2024:PRMFragile}, 
or exhibit the lowering from g- to d-symmetry of NSS~\cite{Belashchenko2024:arxiv}. 
In the effort to verify that a given material is an altermagnet, elucidating their magnetization 
and spin dynamics will provide valuable clues~\cite{Tsymbal:2019}.
In this work, we offer experimental fingerprints to probe 
the emerging class of AFM and altermagnets featuring NSS
by analyzing the transformation of the spin dynamics anisotropy 
in the in-plane applied magnetic field.

In our model, the dynamics of $\bm{S} \parallel \bm{l}$
is driven by $\bm{\Omega} \perp \bm{l}$ 
due to the itinerant electron exchange interaction with the in-plane magnetization. 
Similarly to DMS, 
the exchange interaction enhances an effective electron g-factor, i.e. the giant Zeeman effect,  and allows $\hbar \Omega$ to achieve meV for $B$ of several tesla, making the regimes (i)-(iv) experimentally attainable for moderate $B$. 
Taking $\Omega_{\rm d} =2$~meV, $\tau_2 = 0.25$~ps ($\Omega_{\rm d} \tau_2/2 \approx 0.37$) and $\Omega = 30$~$\mu$eV, we estimate $1/\Gamma \approx 0.9$~ps and 
$1/\Gamma_z^{\prime} \approx 0.55$~ns from Eq.~(\ref{eq:GamZ}); 
the clean limit gives $1/\Gamma_z^\prime = \Omega_{\rm d} / \Omega^2 \approx 1.5$~ns. 
The regime (iv) is achieved for $\Omega \gtrsim 4$~meV at $\tau_2 = 0.25$~ps ($\Omega \tau_2 \gtrsim 1.5$), with $\Gamma_{\pm} \approx 5.8$ ps instead of $0.9$~ps from regime (i). 
The slow dynamics of the longitudinal spin component, $\bm{S} \parallel \bm{l}$, can be additionally affected by the electron-magnon scattering at the elevated temperatures, when the equilibrium population of magnons is significant. In AFM, magnons typically have an energy gap (usually due to magnetic anisotropies) in sub-meV range, which partially decouples them from low-energy electron spin dynamics due to the inelastic scattering. 
Furthermore, the inclusion of SOC might result in the spin-momentum locking hot spots~\cite{Smejkal2022:PRX} along high-symmetry lines in BZ with zero NSS, 
serving as an alternative mechanism of the longitudinal spin relaxation.

The material candidates for altermagnets are diverse~\cite{Yuan2020:PRB,Yuan2021:PRM,Smejkal2022:PRX,Smejkal2022:PRXb,Xiao2023spin,Mazin2021:PNAS,Osumi2024:PRB,Lee2024:Broken,Reimers2024:Direct,duan2024antiferroelectric}, with theoretical estimations of $\Omega_{\rm d}$ varying from few-meV in MnF$_2$~\cite{Yuan2020:PRB,Yuan2021:PRM}, up to tens and hundreds of meV~\cite{Smejkal2022:PRX,Smejkal2022:PRXb,Xiao2023spin}. 
Taking a realistic value of $\tau_2 \approx 0.25$~ps, we expect that the in-plane spin dynamics will display the motion-narrowing induced relaxation for  $\Omega_{\rm d} \lesssim 3$~meV, and will have oscillatory, fast-decaying character in most cases of $\Omega_{\rm d} \gtrsim 3$ meV. 
The regime with a very efficient spin relaxation provides overlooked opportunities for altermagnets, such as imprinting
enhanced spin relaxation in the neighboring semiconductors through proximity effects~\cite{Zutic2019:MT}. 
In contrast to the common goal 
of spintronics to maximize the spin-relaxation time, its reduction to ps and sub-ps range would enable ultrafast switching~\cite{Zutic2004:RMP} 
and a superior operation of  spin-lasers~\cite{Lindemann2019:N} and their electrical control~\cite{Dainone2024:N}.

\acknowledgments
We thank Kirill Belashchenko for helpful discussions.

This work was supported by the Air Force Office of Scientific Research under Award No. FA9550-22-1-0349.

\bibliography{AMR_Ref}

\end{document}